\documentclass{article}

\usepackage{arxiv}
\usepackage[utf8]{inputenc} 
\usepackage[T1]{fontenc}    
\usepackage{url}            
\usepackage{booktabs} 
\usepackage{amsmath}
\usepackage{amsfonts}       
\usepackage{nicefrac}       
\usepackage{microtype}      
\usepackage{physics}
\usepackage{algorithm}
\usepackage{algpseudocode}
\usepackage{graphicx}
\usepackage{doi}
\newtheorem{theorem}{Theorem}[section]

\newtheorem{lemma}[theorem]{Lemma}
\newtheorem{proposition}[theorem]{Proposition}
\usepackage{caption}
\usepackage{subcaption}
\def\bX{\mathbf{X}}

\title{Connecting mass-action models and
network models for infectious diseases}

\author{ Thien-Minh Le\\
  Department of Mathematics\\
 The University of Tennessee at Chattanooga\\
  Chattanooga, Tennessee, U.S.A.\\
  \texttt{thien-le@utc.edu} \\
   \And
 Jukka-Pekka Onnela\\
  Department of Biostatistics\\
  Harvard T.H. Chan School of Public Health\\
  Boston, Massachusetts, U.S.A.\\
  \texttt{onnela@hsph.harvard.edu} \\}

\renewcommand{\shorttitle}{\textit{arXiv} Template}

\begin{document}
\maketitle


\begin{abstract}
Infectious disease modeling is used to forecast epidemics and assess the effectiveness of intervention strategies. Although the core assumption of mass-action models of homogeneously mixed population is often implausible, they are nevertheless routinely used in studying epidemics and provide useful insights. Network models can account for the heterogeneous mixing of populations, which is especially important for studying sexually transmitted diseases. Despite the abundance of research on mass-action and network models, the relationship between them is not well understood. Here, we attempt to bridge the gap by first identifying a spreading rule that results in an exact match between disease spreading on a fully connected network and the classic mass-action models. We then propose a method for mapping epidemic spread on arbitrary networks to a form similar to that of mass-action models. We also provide a theoretical justification for the procedure. Finally, we show the advantages of the proposed methods using synthetic data that is based on an empirical network. These findings help us understand when mass-action models and network models are expected to provide similar results and identify reasons when they do not.
\end{abstract}


\section{Introduction}

Understanding disease spread is crucial for providing accurate predictions of disease outbreaks and for gaining greater insight into prevention strategies. Infectious disease modeling has served as a potent tool for this endeavor for centuries, with the earliest work dating back to Bernoulli (1760). \cite{Bernoulli1760}  Compartmental mass-action models are the most common type of modeling approach, and they are frequently used to study different influenza strains. This modeling assumes that all individuals are well mixed (Kermack, 1927). \cite{Kermack1927} Its advantage is that it is simple to use and has a well-established theoretical foundation for different disease properties, while still producing accurate predictions for multiple types of diseases, particularly influenza. \cite{Nsoesie2014, Afzal2021} Sexually transmitted diseases such as monkeypox, HIV, and HVC are challenging for mass-action models as infected individuals transmit diseases to their neighbors only through a sexual network. Since population structure is naturally represented as a network, there have been numerous investigations into the spread of disease on networks in the last two decades. (\cite{Keeling1997,Newman2002, Kiss2018, Craig2020, Wang2021, Kuga2022}) 

Despite the fact that network epidemiology has received a lot of attention from the research community, most of the work has focused on deriving solutions for spreading processes on static or temporal networks or understanding the effects of network topology characteristics on spreading process outcomes. (\cite{Keeling1997, Newman2002, Kiss2018, Holme2021}) Surprisingly, there are few studies on the connection of network models and mass-action models. Understanding the connection between the two model families is critical because it will allow researchers to see the effect of network topology on the spreading process and could open up new avenues for making use of well-established results from classic models. The first attempt to investigate the relationship between a network model and the classic model was the work of Keeling (2005). \cite{Keeling2005} In this study, the author proposed a modified mass-action model to fit the network model's predictions. The transmission rate of the modified mass-action model was defined to be a function of certain network characteristics (the average degree and the ratio of triangles to triples). Recently, Malloy et al. (2021) used different simulation settings to investigate the influence of the mass-action model and the network model on the effectiveness of prevention strategies. \cite{Malloy2021} 

The purpose of this work is to bridge the gap in the literature by relating the two models. The primary distinction between mass-action models and network-based models lies in their graph structures, as the implicit contact graphs of classic mass-action models are always fully connected, whereas the graphs of network-based models are usually not. In order to connect these two models, we will first study the behavior of epidemic spread on fully connected graphs. There are several ways to define the spreading process on networks, including the Gillespie method, degree infectivity, and unit infectivity methods. (\cite{Kiss2018, Newman2010, Dutta2018}) The degree infectivity method is likely the most prevalent technique, in which each infected node has a fixed probability of transmitting the disease to each of its susceptible neighbors at each time step. To connect the two models, the spreading process on fully connected graphs should yield identical results, regardless of how the spreading rule is defined. Under the degree infectivity method, however, the number of infections on fully connected graphs is always less than the number of infections under the mass-action model. (\cite{Kiss2018}) We first propose a rule for network propagation that eliminates this bias. Then, based on the proposed spreading rule, we present approaches to employ network topology to adapt the mass-action model to capture the spread on networks. We also provide theoretical justifications to support our method. Finally, using simulation and synthetic data, we show the merits of the proposed method in studying epidemics on networks. 

The structure of the paper is as follows. In Section \ref{method}, we discuss the classic mass-action models and our proposed spreading process on networks. Section \ref{proposed} provides the approximation procedure for the proposed spreading process and offers theoretical justifications for it. Section \ref{early behavior} compares the early behavior of the epidemic using the network model and mass-action model. Section \ref{dataanalysis} focuses on data analysis, specifically highlighting the benefits of employing the proposed methods for analyzing epidemics on networks. Finally, Section \ref{discussion} discusses our contribution and possible directions for future research.

\section{Mass-action models and network models}
\label{method}

\subsection{Mass-action models}
Mass action models are the most common model types used in infectious disease epidemiology due to their simplicity. The fundamental assumption of the model is the homogeneous mixing of all individuals, i.e., the contact pattern of individuals forms a fully connected graph. This section examines the SI, SIR, and SITAD processes. Whereas the first two processes, SI and SIR, are frequently used for influenza, for HIV/AIDS we consider the SITAD model, a simplified version of the model used in Hove et al. 2010.\cite{Hove2010}

\subsubsection{SI process}
In the SI process, at a given time, the population is divided into two mutually disjoint compartments: susceptible and infected. Suppose $N$ is the size of the population. Let $S$ and $I$ denote the number of susceptible and infected individuals, respectively ($S+ I = N$). Suppose that the model parameter is $\bold{\theta} = \beta$, where $\beta$ is the transmission rate. Its dynamic states evolve as in Figure \ref{epimodels}a.

 Let us denote the status of its population at time $t$ is $\bX_t = [S_t, I_t]$. Using the tau leaping method by Gillespie (2001) \cite{Gillespie2001}, the status of its population at time $(t  + \tau)$ evolves as $\bX_{t+\tau} = \bX_t +  Y_1 \big(h_1(\bX_t)\tau\big) \nu_1 $, where $\nu_1 = [-1,1]^T$ is the transition vector and $Y_1 \big(h_1(\bX_t)\tau\big)$ is a random variable. Here $Y_1 \big(h_1(\bX_t)\tau\big)$ is Poisson distributed with rate $h_1(\bX_t) \tau = \beta \tau \frac{S_t I_t}{N}$. By choosing $\tau = 1$, which represents the change in population status after each time unit, the dynamic epidemic in the population evolves from $\bX_t = [S_t, I_t]$ to $\bX_{t+1} = [S_{t+1}, I_{t+1}]$ by the transformation $\bX_{t+1} = \bX_t + Y_1 \big(h_1(\bX_t)\big) \nu_1 $. In particular, 
 $ S_{t+1}= S_t - Y_{1,t}$, $I_{t+1} = I_t + Y_{1,t} $, 
where $Y_{1,t}$ is Poisson distributed with rate $h_1(\bX_t) = \beta  \frac{S_t I_t}{N}$.

\subsubsection{SIR process}
The population for the SIR process at a given time is divided into three mutually exclusive compartments: susceptible, infected, and recovered. Suppose $N$ is the size of the population. Let $S, I, R$ denote the number of susceptible, infected, and recovered individuals, respectively ($S + I + R = N$). Suppose that the model parameter is $\bold{\theta} = (\beta, \gamma)$, where $\beta$ is the transmission rate and $\gamma$ is the recovery rate. Its dynamic states evolve as in Figure \ref{epimodels}b.

 Denote the status of its population at time $t$ is $\bX_t = [S_t, I_t, R_t]$. The tau leaping method by Gillespie (2001) \cite{Gillespie2001} tells us the status of its population at time $(t  + \tau)$ evolved as $\bX_{t+\tau} = \bX_t +  \sum_{j=1}^2 Y_j \big(h_j(\bX_t)\tau\big) \nu_j$, where $\nu_1 = [-1,1,0]^T$ and $\nu_2 = [0,-1,1]^T$ are the transition vectors, and $Y_j \big(h_j(\bX_t)\tau\big)$, for $j = 1,2$, are random variables. Here $Y_j \big(h_j(\bX_t)\tau\big)$ Poisson distributed with rates $h_j(\bX_t) \tau$, where $h_1(\bX_t \tau) = \beta \tau \frac{S_t I_t}{N}$ and $h_2(\bX_t \tau) = \gamma \tau I_t$. Let $\tau = 1$, which represents the change in population status after each time unit, the dynamic epidemic in the population evolves from $\bX_t = [S_t, I_t, R_t]$ to $\bX_{t+1} = [S_{t+1}, I_{t+1}, R_{t+1}]$ by the transformation $\bX_{t+1} = \bX_t + \sum_{j=1}^2 Y_j \big(h_j(\bX_t)\tau\big) \nu_j$. In particular, 
 $S_{t+1}= S_t - Y_{1,t}, I_{t+1} = I_t + Y_{1,t} -  Y_{2,t}, R_{t+1}= R_t + Y_{2,t}$, 
where $Y_{1,t}$, $Y_{2,t}$ are Poisson distributed with rates $h_1(\bX_t) = \beta  \frac{S_t I_t}{N}$, $h_2(\bX_t) = \gamma I_t$, respectively.

\subsubsection{SITAD process}
\label{subsec:SITAD mass-action}
\begin{figure}[h]
     \centering
     \includegraphics[width=\textwidth]{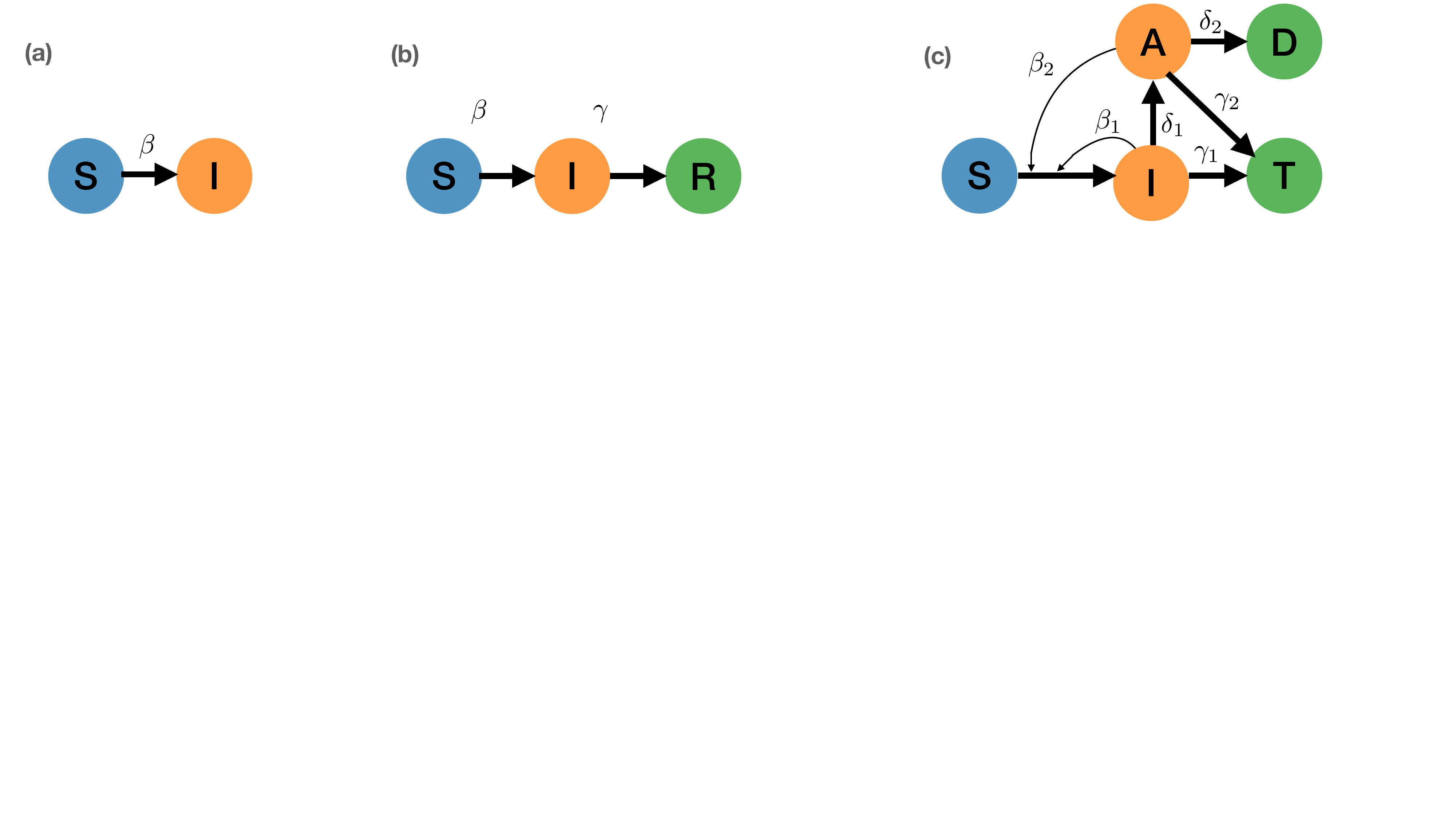}
     \vspace{-7cm}
        \caption{Three different spreading processes: (a) the SI spreading process, (b) the SIR spreading process, and (c) the SITAD spreading process.}
        \label{epimodels}
\end{figure}

We consider a simplified version of the HIV/AIDS model of Hove-Musekwa et al. (2010). \cite{Hove2010} In this model, at a given time, the population's state is divided into five mutually exclusive compartments: susceptible ($S$), HIV positive ($I$), AIDS ($A$), treated ($T$), and deceased ($D$) ($S + I + A + T + D = N$). Its dynamic state evolves as in Figure \ref{epimodels}c, where the model parameter
 $\bold{\theta} = (\beta_1, \beta_2, \gamma_1, \delta_1, \gamma_2 ,\delta_2)$, $\beta_1$ is the transmission rate of HIV, $\beta_2$ is the transmission rate of AIDS, $\gamma_1$ is the treatment rate of HIV, $\delta_1$ is the AIDS progression rate of HIV, $\gamma_2$ is the treatment rate of AIDS, and $\delta_2$ is the death rate of AIDS. 
          
Let the status of its population at time $t$ be $\bX_t = [S_t, I_t, T_t, A_t, D_t]$. Using the tau leaping method by Gillespie (2001) \cite{Gillespie2001}, the dynamic epidemic of the population at time step $(t  + \tau)$ evolves as $\bX_{t+\tau} = \bX_t +  \sum_{j=1}^5 Y_j \big(h_j(\bX_t)\tau\big) \nu_j$, where $\nu_1 = [-1,1,0,0,0]^T$, $\nu_2 = [0,-1,1,0,0]^T$, $\nu_3 = [0,-1,0,1,0]^T$, $\nu_4 = [0,0,1,-1,0]^T$, and $\nu_5 = [0,0,0,-1,1]^T$ are the transition vectors. And $Y_{j,t}$ are Poisson distributed with rates $h_j(\bX_t )$, for $j = 1, \ldots, 5$. Here, $h_1(\bX_t ) = \big(\beta_1 I_t + \beta_2 A_{t} \big) \tau S_t /N, h_2(\bX_t ) = \gamma_1  \tau  I_t,  h_3(\bX_t ) = \delta_1 
 \tau   I_t, h_4(\bX_t ) = \gamma_2  \tau  A_{t}, h_5(\bX_t ) = \delta_2 \tau  A_{t}$. Let $\tau = 1$, which represents the change in population status after each time unit, the dynamic epidemic in the population evolves from $\bX_t = [S_t, I_t, T_t, A_t, D_t]$ to $\bX_{t+1} = [S_{t+1}, I_{t+1}, T_{t+1}, A_{t+1}, D_{t+1}]$ by the transformation $\bX_{t+1} = \bX_t + \sum_{j=1}^5 Y_j \big(h_j(\bX_t)\tau\big) \nu_j$. In particular, 
 $ S_{t+1} = S_t - Y_{1,t}, I_{t+1} = I_t + Y_{1,t} - Y_{2,t} - Y_{3,t}, T_{t+1} = T_{t} + Y_{2,t} + Y_{4,t}, A_{t+1} = A_{t} + Y_{3,t} - Y_{4,t} - Y_{5,t}, D_{t+1} = D_{t} + Y_{5,t},$, 
where $Y_{j,t}$, for $j = 1, \ldots, 5$, are Poisson distributed with rates $h_j(\bX_t )$ that satisfy $h_1(\bX_t ) = \big(\beta_1 I_t + \beta_2 A_{t} \big) S_t /N, h_2(\bX_t ) = \gamma_1   I_t,  h_3(\bX_t ) = \delta_1 
   I_t, h_4(\bX_t ) = \gamma_2    A_{t},$ and $ h_5(\bX_t ) = \delta_2  A_{t}$.

\subsection{Network models and the proposed spreading process}

The graphs of the network-based model are usually not fully connected, so the spreading process depends on network topology. For simplicity, we consider a given fixed network $G$ with a known initial single infected node. There are several methods to define the spreading process on networks, and the behavior will depend on the method used. The degree infectivity method, the unit infectivity method, and the Gillespie spreading method are some examples of popular spreading methods. (\cite{Newman2010, Kiss2018, Dutta2018}) Among these spreading rules, the degree infectivity method is probably most common as it is straightforward and allows the study of different epidemic behavior, such as the reproductive numbers $R_0$ and $R_*$. \cite{Volz2009} In the degree infectivity method, spreading occurs between infected and susceptible individuals, where at each time step each infected node has a fixed probability of transmitting the disease to each of its susceptible neighbors. Note that the meaning of the transmission parameter under degree infectivity differs from that of the transmission parameter in the mass-action model. In the network model, the transmission parameter represents the transmission rate from each infected individual to each of its neighbors, whereas the transmission rate in the mass-action model specifies the transmission rate from each infected individual to the whole network. Therefore, on a graph with $N$ fully connected nodes, the transmission parameter $\beta$ of the mass-action model corresponds to the transmission rate $\beta/N$ in the network model.

To evaluate the link between the network model and the traditional mass-action model, we first investigate the epidemic behavior of both models when the graph is fully connected. Because the topologies of the populations under the two models are identical, we expect that their behaviors will be identical. Under degree infectivity, however, the network model on a fully connected graph results in fewer infections than the mass-action model.\cite{ Kiss2018} The gap is intuitively caused by the fact that the network model generates new infections at the local level and then aggregates them at the global level, whereas the mass-action model generates new infections at the global level and then allocates them randomly to local positions. Therefore, the network model may produce a lower number of infections when, at the local level, one susceptible node may be infected by two or more of its infected neighbors. This underestimation is more problematic if the number of infected nodes in a network exceeds the number of susceptible nodes, causing infected nodes in a sense to ``compete'' for susceptible nodes.  Therefore, as long as this bias persists, the gap between the two models persists. We first propose a spreading rule on fully connected networks that allows for an exact match with the number of infections.

Under the proposed spreading rule, at each time step, the total transmission rate is computed first, and the number of new infections is then generated based on the total transmission rate. New infections are assigned to at-risk nodes (susceptible neighbors of infected nodes) using a weighted random sample, where the weight of each at-risk node is proportional to the number of infected neighbors. In particular, let $\beta$ denote the transmission rate of each infected node to the whole network in the mass-action model, $\bold{I}_t$ is the set of infected nodes at time $t$, $S_{i,t}$  the number of susceptible neighbors of node $i$ at time $t$, $k_i$  the number of neighbors of node $i$, and $S^*_t$ the number of at-risk nodes at time $t$. The total transmission rate at time $t+1$ is calculated as $ \beta  \sum_{i \in \bold{I}(t)}\dfrac{S_{i,t}}{k_i+1}$. The number of new infections $Y_{1,t}$ is generated from $ \text{Binom} \big(S^*_t, h_1(\bX_t )/S^*_t \big)$, and the new infections are randomly allocated among the at-risk nodes based on their weights (see algorithm \ref{SIproposed} in the Appendix for more details). Under the proposed spreading process, when the network is fully connected, susceptible nodes and at-risk nodes are the same, i.e., $S^*_t = S_t$. Therefore, at time $t$, the total transmission rate is $\beta I_t S_t/N $, and new infections are generated from Binomial$(S_t, \beta I_t/N   )$. Our proposed procedure allows for an exact match with the mass-action model. 

\begin{figure}[h]
	\centering
	 \includegraphics[width=\textwidth]{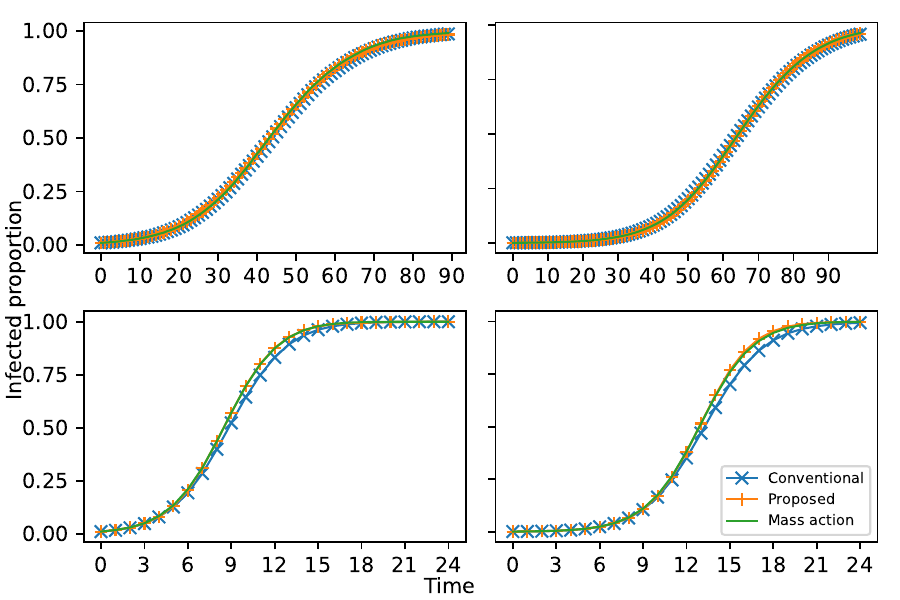}
  
	\caption{Comparison of the proposed network spreading rule (Proposed) on fully connected graphs, the conventional network degree infectivity spreading rule (Conventional), and the mass-action SI model (Mass-action) with the transmission parameter $\beta$ = 0.12 with 100 nodes (top left) and 1000 nodes (top right), and $\beta$ = 0.7 on fully connected graphs with 100 nodes (bottom left) and 1000 nodes (bottom right).}
	\label{fig:Agreementcompletegraph}
\end{figure}

Figure \ref{fig:Agreementcompletegraph} displays the average proportion of infections over time for our spreading rule, the degree infectivity spreading rule, and the mass-action model on fully connected graphs. Here we consider four cases: graphs of 100 nodes and 1000 nodes with the transmission parameter $\beta = 0.12$ (top left and top right), and graphs of 100 nodes and 1000 nodes with the transmission parameter $\beta = 0.7$ (bottom left and bottom right). The average is taken over 200 stochastic realizations. As expected, as the transmission parameter is small, the different spreading rules are hard to distinguish. But when the transmission parameter is large, the proposed spreading rule still precisely matches the mass-action model, but the degree infectivity spreading rule on the network underestimates the number of infections relative to the mass-action model. We also notice that the bias is unaffected by the size of the network. The proposed spreading rule therefore removes the bias.

\section{Approximations to the proposed spreading process}\label{proposed}

The most straightforward strategy for studying epidemic spread on network is to use simulation. Despite the fact that this method provides us with many insights into various disease-spreading processes in a variety of settings, it does not give us a thorough theoretical understanding of the spreading process. Much effort has been dedicated to investigating the solution of the spreading process using the degree infectivity rule. The bottom-up approach is the most common where nodes are evaluated in pairs, which can then be evaluated in triples, and so on. Some strategies for approximating higher order structure by lower order structure have been proposed, such as approximating triples by pairs. Finally, we can obtain the average number of nodes in each state by solving a system of differential equations and using node and pair relations (for more details, see \cite{Kiss2018}). Although these approaches provide alternatives to simulating the network spreading process, they are local approaches as they do not account for the entire network topology. In addition, there is no closed form solution and instead results must be evaluated numerically. In this sense, these approaches provide limited insight into the relationship between network models and mass-action models.

In the following, we present another approach for a better understanding of the relationship between network models and mass-action models. The main idea is to set up a system of equations that are analogous to those of the mass-action model while also taking into account the topology of the network. For example, for the SI process, we aim to replace $\beta I S/N$ with $I \beta(G)$, where $\beta(G)$ is a function that contains information about network topology.

\subsection{The modified SI process }

\begin{figure}
     \centering
         \includegraphics[width=\textwidth]{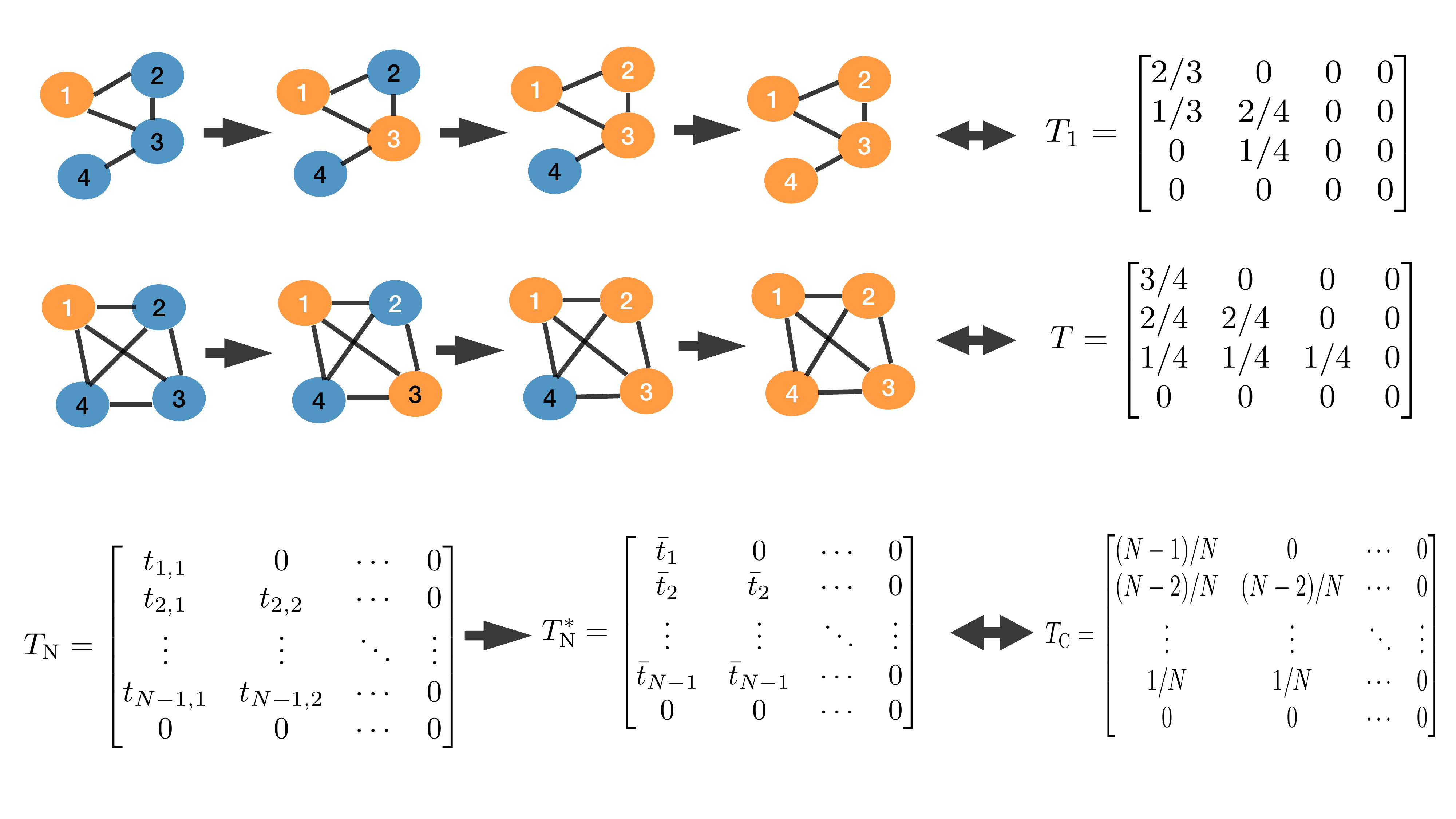}
      \centering
         \caption{The connection in transmission rate between network model and mass-action model. (a) Transmission order on a network of four nodes and its corresponding transmission matrix. (b) Transmission order on a complete network of four nodes and its corresponding transmission matrix. (c) Connection of the network transmission matrix and the transmission matrix corresponding to the mass-action model. }
          \label{Transmission-matrix-construction}
\end{figure}

Different nodes are infected at different times in the course of the SI process. If the order of infections is known, we introduce a transmission matrix $\bold{T} = [\bold{T}_{i,j}]_{N \times N}$ based on network topology. Element $(i,j)$ of the transmission matrix $\bold{T}$, $\bold{T}_{i,j}$, represents the transmission rate of node $j$ when the network has $i$ infected nodes. Here $\bold{T}_{i,j} = S_{i,j}/(k_j+1)$, where $S_{i,j}$ is the number of susceptible neighbors of node $j$ when there are $i$ infected nodes in the network, and $k_j$ is the degree of node $j$. Therefore, the summation of row $i$ in the transmission matrix,$\sum_{j=1,N}\bold{T}_{i,j}$, gives the overall transmission rate when there are $i$ infected nodes in the network. Since reordering the column of the transmission matrix $\bold{T}$ will not change the overall transmission rate (the row sum), for simplicity, we rearrange the columns of the transmission matrix in the order of infections and use this rearranged matrix as our transmission matrix $\bold{T}$. Figure \ref{Transmission-matrix-construction} demonstrates the transmission matrices corresponding to the infection order (1,3,2,4), where these matrices result from reordering the columns of the original transmission matrices from (1,2,3,4) to (1,3,2,4) as \begin{equation*}
\bold{T}_1 = 
\begin{bmatrix}
2/3 & 0 & 0 & 0 \\
1/3 & 0 & 2/4 & 0 \\
0 & 0  & 1/4 & 0 \\
0 & 0 & 0&0 
\end{bmatrix} \rightarrow \bold{T}_1 = 
\begin{bmatrix}
2/3 & 0 & 0 & 0 \\
1/3 & 2/4 & 0 & 0 \\
0 & 1/4  & 0 & 0 \\
0 & 0 & 0&0 
\end{bmatrix} \text{and} 
\end{equation*} 

\begin{equation*}
 \bold{T}_c = 
\begin{bmatrix}
3/4 & 0 & 0 & 0 \\
2/4 & 0 & 2/4 & 0 \\
1/4 & 1/4  & 1/4 & 0 \\
0 & 0 & 0&0 
\end{bmatrix} \rightarrow \bold{T}_c =  \begin{bmatrix}
3/4 & 0 & 0 & 0 \\
2/4 & 2/4 & 0 & 0 \\
1/4 & 1/4  & 1/4 & 0 \\
0 & 0 & 0&0 
\end{bmatrix} .
\end{equation*} 

Let us denote $\sum_{k=1}^N \bold{T}[I_t,k] = I_t \bar{T}_{I_t}$. The behavior of the spreading process on the network using the transmission matrix $\bold{T}$ after each time unit then can be described by \begin{equation}
\label{SIequationnetwork}
    \bX_{t+1} = \bX_t + Y_1 \big(h_1(\bX_t)\big) \nu_1, 
\end{equation}

where $ \bX_{t+1} = (S_{t+1}, I_{t+1}), \bX_{t} = (S_{t}, I_{t}), \nu_1 = [-1,1]^T$, and $Y_{1,t}$ is Poisson distributed with rate  $h_1(\bX_t )$ = $\beta  I_t \bar{T}_{I_t}$.






Compared to the SI mass-action model, it is apparent that the SI process on networks is controlled by network topology through the transmission matrix $\bold{T}$, where the mass-action transmission rate $\beta I_t S_t/N $ is replaced by $\beta I_t \bar{T}_{I_t}$. Therefore, the transmission matrix $\bold{T}$ contains the entirety of network information. The relationship between a network model and the mass-action model is depicted in Figure \ref{Transmission-matrix-construction}c. In contrast to the transmission matrix of the conventional mass-action model, which is always $\bold{T_C}$, the transmission matrix of the network model will vary according to the network topology and takes the form $\bold{T_N}$. Under our procedure, the actual transmission matrix $\bold{T_N}$ is approximated by $\bold{T^*_N}$, where non-zero elements in each row represent the average infection rate at that time.

Let ${I_{\bold{T},t}}$ denote the sequence of the number of infected nodes based on the transmission matrix $\bold{T}$ at time  $t$. Denote $E(  I_{\bold{T},1}^+ ) = E\big( I_{\bold{T},1}\vert I_{\bold{T},0} = 1 \big) $, $\cdots,$ $E(  I_{\bold{T},k}^+ ) = E\big( I_{\bold{T},k}\vert I_{\bold{T},k-1} = E(  I_{\bold{T},k-1}^+ )\big) $. Therefore, the average number of infected nodes from time 1 to $K$ in the network is $E(  I_1^+ ), \cdots, E(  I_K^+)$. Similarly, the average number of susceptible nodes from time 1 to $K$ in the network is $E(  S_1^+ ), \cdots, E(  S_K^+)$.

Lemma \ref{SIlemma} below tells us that the proposed network spreading process and the modified network spreading process have the same average realization.

\begin{lemma}
\label{SIlemma}
When the order of infections is known, the spreading process based on equation (\ref{SIequationnetwork}) has the same average realization as the proposed SI process on the network.
\end{lemma}

If the order of infections is unknown, we can obtain the infection order by a random sample. In this approach, each newly infected node is incrementally updated by sampling at-risk nodes based on their risk weights (see Algorithm \ref{samplingorder} in the Appendix for details). Then, using the same procedure as before, we can construct the corresponding transmission matrix for each sample and calculate the number of infections. Finally, the average number of infections is obtained by averaging the number of infections corresponding to the sampled infection order sequences. This sampling scheme has the following rationale. Consider the network $G$ with $N$ nodes and a known first infected node; there are at most $(N-1)!$ possible infection order sequences. Assuming that we obtained $m$ infection order sequences using the sampling approach, let $\bold{T}_i$ denote the transmission matrix corresponding to the infection order sequence $i$. The average realization of the number of infections can then be approximated by the average realization of the number of infections from transmission matrices $\bold{T}_i$. Therefore, as the sample size $m$ grows, the average number of infections based on transmission matrices $\bold{T}_i$ will converge to the average realization of the spreading process on the network.  

\textit{Remark 1:} For some networks, the matrices constructed as described above do not change. Complete networks, k-star networks, and cycle networks are all examples of these kinds of networks.

\subsection{The modified SIR process}
\label{SIRproposed}
Similarly to the SI process, we first consider the infection order of all nodes when it is known, and when it is unknown, we use the same sampling procedure as above. Unlike in the SI process, when a node recovers in the SIR process, the total number of at-risk nodes changes. Due to network topology, the same number of recovered nodes may result in a different number of at-risk nodes. If recovery occurs at random, the Binomial approximation cannot be used. Therefore, the exact match in terms of the number of infections will only be feasible if the recovery order is determined by the length of time a node was infected. As random recovery is a common assumption, we will focus on this case. As the number of at-risk nodes is unattainable, we can generate the number of newly infected nodes at each time step using the Poisson distribution. For a given infection order sequence, we can extract the transmission matrix $\bold{T}$ corresponding to the case in which there is no recovery during transmission.

Denote $ \bar{T}_{H} = \sum_{k=1}^N \bold{T}[H,k]/H, H = I + R$. The modified SIR spreading process based on the transmission matrix $\bold{T}$ can now be updated as  \begin{equation}
\label{SIRequationnetwork}
    \bX_{t+1} = \bX_t + \sum_{j=1}^2 Y_j \big(h_j(\bX(t))\tau\big) \nu_j, 
\end{equation}

where $ \bX_{t+1} = (S_{t+1}, I_{t+1}, R_{t+1}), \bX_{t} = (S_{t}, I_{t}, R_{t}), \nu_1 = [-1,1,0]^T, \nu_2 = [0,-1,1]^T$; $Y_{1,t}$ and $Y_{2,t}$ are Poisson distributed with rates $h_1(\bX_t) =   \beta I_t \bar{T}_{H}$ and $h_2(\bX_t) = \gamma I_t$, respectively.



Let ${I_{\bold{T},t}}$ and ${R_{\bold{T},t}}$ denote the sequence of the number of infected nodes and recovered nodes based on the transmission matrix $\bold{T}$ at time $t$, respectively. Denote $E(  I_{\bold{T},1}^+ ) = E\big( I_{\bold{T},1}\vert I_{\bold{T},0} = 1, R_{\bold{T},0} = 0 \big) $, $E(  R_{\bold{T},1}^+ ) =  E\big( R_{\bold{T},1}\vert I_{\bold{T},0} = 1, R_{\bold{T},0} = 0 \big) $, $\cdots,$ $E(  I_{\bold{T},k}^+ ) = E\big( I_{\bold{T},k}\vert I_{\bold{T},k-1} = E(  I_{\bold{T},k-1}^+ ), R_{\bold{T},k-1} =  E(  R_{\bold{T},k-1}^+)  \big) $, $E(  R_{\bold{T},k}^+ ) = E\big( R_{\bold{T},k}\vert I_{\bold{T},k-1} = E\big(  I_{\bold{T},k-1}^+ \big), R_{\bold{T},k-1} = E(  R_{\bold{T},k-1}^+ ) \big)$. Then the average realization of the number of nodes in each state from time 1 to $K$ on the network is $E(  Y_1^+ ), \cdots, E(  Y_K^+),$ where $Y$ is one of the states $S, I, R$. 

The following lemma shows that the two spreading processes have the same average realization. 

\begin{lemma}
\label{SIRlemma}
  If the infection order sequence is known, the spreading process based on equation (\ref{SIRequationnetwork}) has the same average realization as the SIR process on the network. 
\end{lemma}


\subsection{The modified SITAD process}

The SITAD process on networks starts from an initially infected node and then spreads to cause new HIV infections with rate $\beta_1$. Among those infected with HIV, some progress to AIDS and some get treated. Individuals with AIDS spread the disease and cause new HIV infections with rate $\beta_2$. Among individuals with AIDS, some will get treated and some will die. The risk weight of each at-risk node in the SITAD spreading process is determined by $w = \beta_1 \times \text{number HIV neighbors} + \beta_2 \times \text{number AIDS neighbors}$. Similarly to the SIR process, we consider the case where AIDS progression, treatment, and death happen at random, and the infection order is known. Let $\bold{T}$ be the transmission matrix corresponding to the infection order sequence. The modified SITAD spreading process based on the transmission matrix $\bold{T}$ can now be updated as follows:

Denote $ \bar{T}_{H} = \sum_{k=1}^N \bold{T}[H,k]/H, H = I +  T + A + D$. Here $H$ is the total number of people with HIV. The modified SITAD spreading process based on the transmission matrix $\bold{T}$ can now be updated as  \begin{equation}
\label{SITADequationnetwork}
    \bX_{t+1} = \bX_t + \sum_{j=1}^5 Y_j \big(h_j(\bX(t))\tau\big) \nu_j, 
\end{equation}

where $ \bX_{t+1} = [S_{t+1}, I_{t+1}, T_{t+1}, A_{t+1}, D_{t+1}], \bX_t = [S_t, I_t, T_t, A_t, D_t], \nu_1 = [-1,1,0,0,0]^T, \nu_2 = [0,-1,1,0,0]^T, \nu_3 = [0,-1,0,1,0]^T, \nu_4 = [0,0,1,-1,0]^T, \nu_5 = [0,0,0,-1,1]^T$. $Y_{j,t}$, for $j = 1, \ldots, 5$, are Poisson distributed with rates $h_j(\bX_t )$. Here, $h_1(\bX_t ) =(\beta_1  I + \beta_2  A) \bar{T}_{H}, h_2(\bX_t ) = \gamma_1   I_t,  h_3(\bX_t ) = \delta_1 
   I_t, h_4(\bX_t ) = \gamma_2    A_{t},$ and $ h_5(\bX_t ) = \delta_2  A_{t}$.



Similarly as for the SIR process, we define the average realization for the SITAD process based the transmission matrix $\bold{T}$ at time $k$ as $E(  I_{\bold{T},k}^+ ) = E\big( I_{\bold{T},k}\vert I_{\bold{T},k-1} = E(  I_{\bold{T},k-1}^+ ), T_{\bold{T},k-1} =  E(  T_{\bold{T},k-1}^+ ), A_{\bold{T},k-1} = E(  A_{\bold{T},k-1}^+), D_{\bold{T},k-1} =  E(  D_{\bold{T},k-1}^+ )  \big) $, $E(  T_{\bold{T},k}^+) = E\big( T_{\bold{T},k} \vert I_{\bold{T},k-1} = E(  I_{\bold{T},k-1}^+ ), T_{\bold{T},k-1} =  E(  T_{\bold{T},k-1}^+), A_{\bold{T},k-1} = E(  A_{\bold{T},k-1}^+ ), D_{\bold{T},k-1} =  E(  D_{\bold{T},k-1}^+)  \big) $, $E(  A_{\bold{T},k}^+) = E\big( A_{\bold{T},k} \vert I_{\bold{T},k-1} = E(  I_{\bold{T},k-1}^+ ), T_{\bold{T},k-1} =  E(  T_{\bold{T},k-1}^+ ), A_{\bold{T},k-1} = E(  A_{\bold{T},k-1}^+ ), D_{\bold{T},k-1} =  E(  D_{\bold{T},k-1}^+ )  \big) $, $E(  D_{\bold{T},k}^+) = E\big( D_{\bold{T},k} \vert I_{\bold{T},k-1} = E(  I_{\bold{T},k-1}^+ ), T_{\bold{T},k-1} =  E(  T_{\bold{T},k-1}^+ ), A_{\bold{T},k-1} = E(  A_{\bold{T},k-1}^+), D_{\bold{T},k-1} =  E(  D_{\bold{T},k-1}^+ )  \big) $. The average realization of the number of nodes in each state from time 1 to K in the network are $E\big(  Y_1^+ \big), E\big(  Y_2^+ \big), \cdots, E\big(  Y_K^+ \big),$ where $Y$ is one of the states $S, I, T, A, D$.

The following lemma tells us that the average realizations based on the two approaches are the same.

\begin{lemma}
\label{SITADlemma}
  If the order of infections is known, the spreading process based on equation (\ref{SITADequationnetwork}) has the same average realization as the SITAD process on the network. 
\end{lemma}

\subsection{Approximations of the spreading processes using the average transmission matrix}
\label{averagetransmission}

As shown in the preceding sections, when the infection sequence is known, the modified SI, SIR, and SITAD processes generate the same average number of infections on networks. Since the order of infections is often unknown, we can determine the order of infections using random sampling. We proved that when the infection order is known, the average realization of the number of infections based on the corresponding transmission matrix equals the average realization based on the network. As a result, the average number of infections based on transmission matrices $\bold{T}_i,$ for $ i=1,\cdots, m$ will converge to the average realization of the network spreading process as the sample size $m$ grows.

For a given infection sequence with the transmission matrix $\bold{T}_i$, if the transmission rate is $\beta$, $\beta \bold{T}_i$ represents the matrix of average spreading rates corresponding to the infection sequence. Therefore, the average spreading rate corresponding to $m$ different realizations $\bold{T}_i,$ for $ i=1,\cdots, m$ can be approximated by $\beta \bold{\bar{T}}$, where  $ \bold{\bar{T}} = 1/m \sum_{i=1}^m \bold{T}_i$. In other words, the average number of infections based on the modified process utilizing the average transmission matrix $\bold{\bar{T}}$ can be used to approximate the average number of infections generated by the network spreading process. We refer to this approximation approach as the average transmission matrix model, or ATMM. 



\section{The early behavior of the proposed SIR spreading process on networks}
\label{early behavior}
In this section, we investigate the behavior of the reproductive number for the proposed SIR process on networks. From Lemma \ref{SIRlemma}, we know that once the order of infections is known, the proposed SIR spreading process on a network has the same average realization as the modified spreading process based on the transmission matrix corresponding to the infection order as described in the system of equations (\ref{SIRequationnetwork}). Therefore, we can study the basic reproductive number of the SIR process on a network by using the system of equations (\ref{SIRequationnetwork}). 

We have $\dfrac{dI}{dt} =  I  \beta \bar{T}_{H} - \gamma I = I(\beta \bar{T}_{H} - \gamma )$. So the basic reproductive number $R_0$ of the spreading process on the network is $R_0 = \dfrac{\beta}{\gamma} \bar{T}_{1}$, and the epidemic is possible iff $R_0 > 1$. Let us denote the sequence of distinct node degrees of the given network as $\{ k_1 , \cdots, k_s\}$; the probability a given node has degree $k_i$ is $p_i$, where $\sum_{i=1}^s p_i = 1$ and $1 \leq s \leq N$. If the initial infected node is node $i$, then the basic reproductive number $R_0$ is $R_0 = \dfrac{\beta}{\gamma} \bar{T}_{1} = \dfrac{\beta}{\gamma} \dfrac{k_i}{k_i + 1} =  \dfrac{\beta}{\gamma} (1 - \dfrac{1}{k_i + 1})$. 
If the initially infected node is unknown, the basic productive number now follows a distribution induced by the node degree distribution where $R_0 = \dfrac{\beta}{\gamma} (1 - \dfrac{1}{k_i + 1})$ with probability $p_i$, where $p_i$ is again the probability a node has degree $k_i$. The average basic reproductive number is $\bar{R_0}  =  \sum_{i=1}^s p_i \dfrac{\beta}{\gamma} (1 - \dfrac{1}{k_i + 1}) = \dfrac{\beta}{\gamma} ( 1 - \sum_{i=1}^s p_i  \dfrac{1}{k_i + 1})$. 

Denote $k^* = \text{min} \{ k_1 , \cdots, k_s\}$. We have the following bounds: $\dfrac{\beta}{\gamma} ( 1 -  \dfrac{1}{k^* + 1}) < \dfrac{\beta}{\gamma} ( 1 -  \dfrac{1}{k_i + 1}) <  \dfrac{\beta}{\gamma} ( 1 - \dfrac{1}{N})$ for all $i \in {1,\cdots, k}$. Since the network structure underlying the mass action model is a fully connected network, its basic reproductive number is $R_0^c = \dfrac{\beta}{\gamma} ( \dfrac{N-1}{N}) = \dfrac{\beta}{\gamma} ( 1 - \dfrac{1}{N}) $. We see that the quantity $R_0^c $ is an upper bound on the basic productive number of a spreading process on a network. This tells us that if the process starts with one initially infected node, the spreading process on a network will less likely lead to an epidemic compared to the mass-action model. The lower bound of the above inequality tells us that the epidemic will least likely occur if the initially infected node has the smallest number of neighbors (smallest degree). 

Next, we consider the early stage behavior of the effective reproductive number $R_t$, for $t$ small. For simplicity, suppose that at time $t$, the network has $h$ infected nodes and no recovered nodes. Let $\bold{I}$ denote the set of $h$ infected nodes at time $t$ and let $k^*_\bold{I} = \min \{ k_i\}_{i \in \bold{I}}$. We have $R_t = \dfrac{\beta}{\gamma} \bar{T}_{h} = \dfrac{\beta}{ \gamma} \dfrac{1}{h} \sum_{i \in \bold{I}}\dfrac{S_i}{k_i + 1}$. Since $ 1 - \dfrac{h}{k_i + 1} \leq \dfrac{S_i}{k_i + 1} \leq 1 - \dfrac{2}{k_i + 1}$,  the lower bound $R^L_t$ and upper bound $R^U_t$ of $R_t$ are given by $ R^L_t = \dfrac{\beta}{ \gamma} ( 1 - \sum_{i \in \bold{I}}  \dfrac{1}{k_i + 1}) \leq   R_t  \leq R^U_t = \dfrac{\beta}{ \gamma} (1 - \dfrac{2}{h} \sum_{i \in \bold{I}}  \dfrac{1}{k_i + 1})$. We observe that the effective reproductive number of the spreading process on networks attains its upper bound if the spreading path of $h$ infected nodes form a line, and it attains its lower bound if the spreading path of $h$ infected nodes forms a complete graph of $h$ nodes.  

Finally, we consider the important question of whether there are any scenarios where the spreading process on a network is more aggressive than the mass-action model (where its corresponding network structure is fully connected). The following Proposition gives us the answer to this important question. 

\begin{proposition}\label{Rt_agressive}
Consider the SIR process on networks at the early stage with $h$ infected nodes.
\begin{itemize}
    \item[a.] For large networks as size  $N \rightarrow \infty$, the asymptotic behavior of the effective reproductive number $R_t$ at the early stage of the epidemic on networks is always asymptotically bounded above by the effective reproductive number of the mass-action model.
    \item[b.] For finite-size networks, the effective reproductive number $R_t$ at the early stage of the epidemic on networks is greater for the mass-action model if all $h$ infected nodes form a line graph and each infected node has more than $2N/h -1$ susceptible neighbors.
\end{itemize}
   
\end{proposition}

Proposition \ref{Rt_agressive}  shows that for a large network, its effective reproductive number $R_t$ at the early stage ($t$ small) is always asymptotically bounded from above by its counterpart mass-action model. However, given a finite-size network, the network spreading process can be more aggressive than the mass-action model depending on the spreading pattern and network topology. This highlights the importance of network topology in understanding disease dynamics.  

\begin{figure}[h]
     \centering\includegraphics[width=\textwidth]{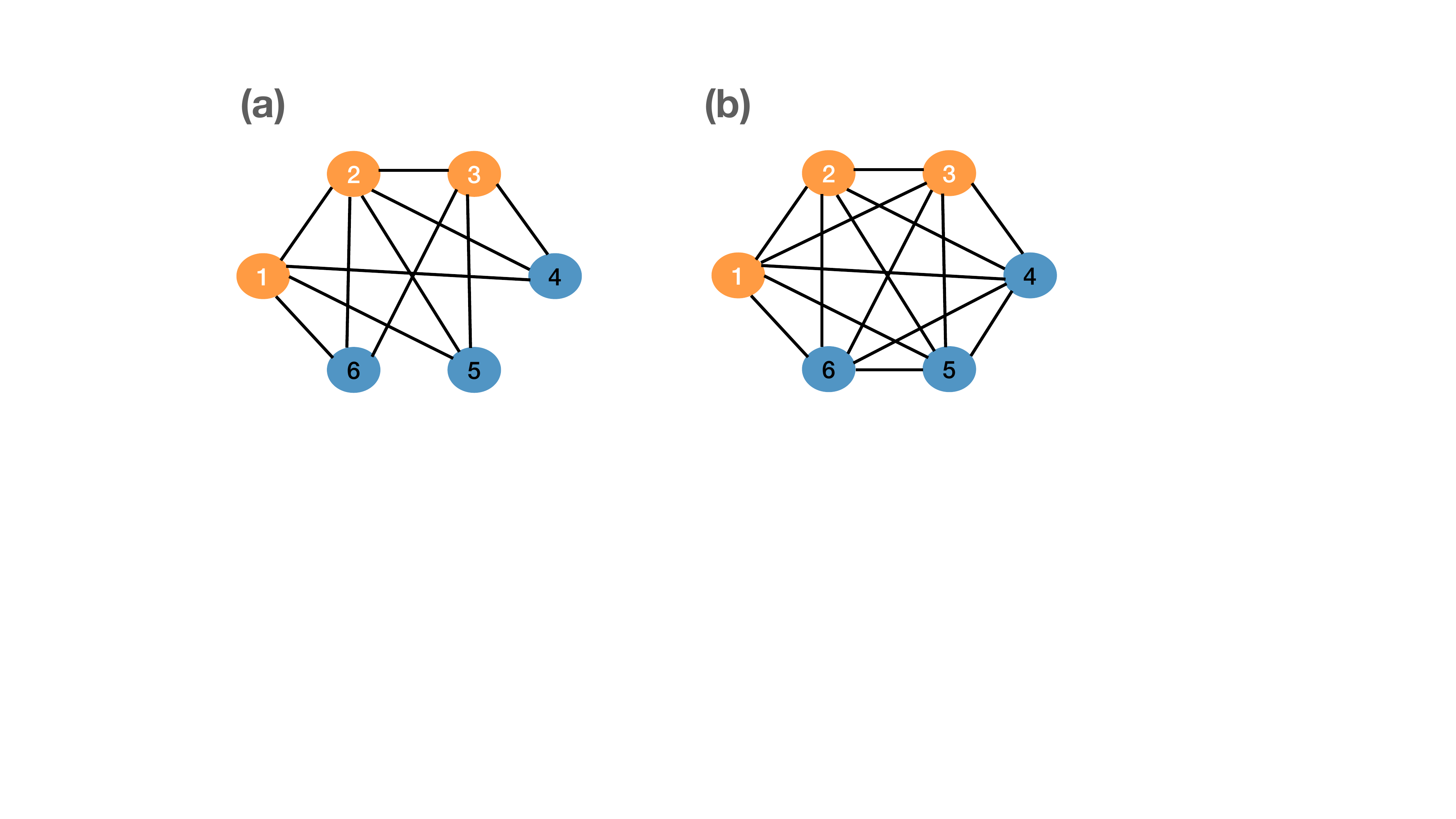}
     \vspace{-5cm}
         \caption{Comparing transmission rates of two networks with three infected nodes (1,2,3): (a) partially connected network with the total transmission rate of 2.1, and (b) fully connected network with the total transmission rate of 1.5 }
         \label{spreadingstatus}
\end{figure}

Figure \ref{spreadingstatus} demonstrates the case where the spreading process on a partially connected network is more aggressive than the spreading process on a fully connected network. In this case, the total transmission rate on the partially connected network is 2.1, while the total transmission rate on the fully connected network is 1.5.


\section{Data Analysis}
\label{dataanalysis}

In this section, we first use simulation to show that the modified spreading process agrees with the proposed network spreading process in terms of the average number of infections over time. We next use synthetic network data to show how the modified spreading process outperforms the proposed spreading process in terms of computation. Finally, we demonstrate that using the network model to analyze network epidemic data, once network information is accessible, surpasses the mass-action model both in terms of computational efficiency and goodness of fit.   

\subsection{Modified spreading spreading processes }
\label{simulation}

\begin{figure}[h]
     \centering\includegraphics[width=\textwidth]{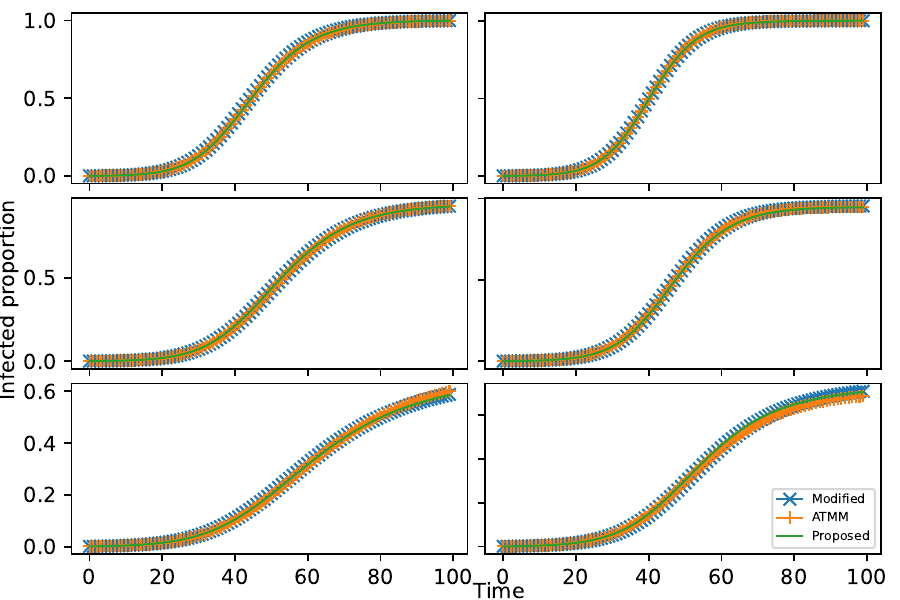}
         \caption{Approximation of different approaches for the SI, SIR, and SITAD spreading processes on the BA network (left) and modified ER network with a single connected component (right). The first row corresponds to the SI process with the model parameter $\bold{\theta} = 0.2$, the second row to the SIR process with $\bold{\theta} =  (0.18, 0.01)$, and the last row to the SITAD process with $\bold{\theta} = (0.2,0.1,0.05,0.03,0.02,0.01)$.  }
         \label{approximate}
\end{figure}

We conducted simulation studies for the three processes discussed in the paper: SI, SIR, and SITAD. For each process, the fixed network structure was generated from a network model. We considered networks of $N = 1000$ nodes generated by the Erdos-Renyi (ER) model with the probability parameter $p = 0.1$, and the Barab\'asi–Albert (BA) model with the parameter 15. Since the ER network can have multiple components, we forced it to one component by adding the set of edges $\{(1,2), (2,3), \cdots, (N-1, N) \} $. Without loss of generality, we assumed that node 1 is the initial infected node. Based on the given network, the initially infected node, and the model parameter $\bold{\theta}$, the average number of infections is determined by averaging the number of infections resulting from 1000 iterations of disease transmission using the proposed spreading rule on the network. On the other hand, using the sampling procedure, we generated 30 infection order sequences and their corresponding transmission matrices $\bold{T}_i$, for $i=1,\cdots,30$. Then, we used the modified process that utilized random transmission matrices $\{\bold{T}_i\}_{i=1,\cdots,30}$ to produce 1000 realizations of the number of infections. We obtained the average realization of the number of infections of the modified process by taking the average of these infection sequences. We also considered the ATMM by applying the modified process to the average transmission matrix $\bold{\bar{T}} = \frac{1}{30}\sum_{i=1}^{30} \bold{T}_i$. In particular, we simulated the modified process utilizing the average transmission matrix 1000 times. We then calculated the average number of infections by averaging those 1000  simulated realizations. Figure \ref{approximate} shows a good agreement across the different approaches.

In Sections \ref{ModifiedSIRvsProposedSIR} and \ref{ModifiedSIRVs.Mass-actionSIR}, we use synthetic network data to illustrate the benefits of the modified spreading process in estimating model parameters. The synthetic data is generated using the proposed discrete time SIR process in Section \ref{SIRproposed} on an empirical network dataset. The empirical network data is an aggregate of network data obtained from the Copenhagen Network Study (CNS), which was made publicly accessible in 2019 .\cite{sapiezynski2019interaction} The network data  comprises the connectivity patterns of 706 students at the Technical University of Denmark during a 28-day period in February 2014. The connectivity patterns are identified through the use of Bluetooth as participants consented to use loaner phones provided by the study as their main phone throughout the study. The received signal strength indicator (RSSI), which can serve as an approximation of physical distance, was collected every five minutes. Following Hambridge et al. in \cite{hambridge2021}, we assigned a connection between two persons if there was at least one RSSI signal large enough during the period, i.e., RSSI $\geq \text{-75dBm}$. For analysis purposes, we simply kept the largest component, which contained 673 nodes and 57,712 edges, as a fixed network. Based on the fixed empirical network, we synthesized epidemic data. For generating network epidemic data, we used the discrete time SIR proposed spreading process as described in Section \ref{SIRproposed}. In particular, we first generated model parameters from prior distributions and then simulated network epidemic data using the generated parameters. If the synthesized data realization was good enough, meaning there was enough data to estimate model parameters, we kept it and retained the model parameters. Since there were only 673 nodes in the observed network, we specified that a good realization had to have cumulatively at least 50\% of nodes infected and 10\% of nodes recovered. Once these constraints were met, the synthesized data was treated as observed network epidemic data, with the corresponding parameters serving as the underlying truth to evaluate the accuracy of estimation.


\subsection{Proposed SIR Model and the SIR ATMM}\label{ModifiedSIRvsProposedSIR}

This section presents a comparison of the performance of the proposed SIR process and the SIR ATMM in estimating model parameters based on observed network data. Due to the intractability of the likelihood function for the epidemic on networks, we employ approximate Bayesian computation (ABC), a method that does not rely on likelihood to estimate the parameters of the model. The variant ABC method used in this paper is replenishment ABC (RABC). \cite{Drovandi} Criteria for comparison include computational time, confidence interval coverage rate from posteriors, and interquartile range. To obtain these metrics, we setup the code as follows.

\textit{Step 1. Generating data and parameters}: For $i \in \{1,\cdots,100\}$, we generate the parameter $\theta^{(i)} = (\beta^{(i)},  \gamma^{(i)}) $ from uniform priors $\beta^{(i)} \sim U(0, .3) $ and $\gamma^{(i)} \sim U(0, .2) $. Based on the parameters, empirical network, and the proposed SIR model, we generate a data set $\text{Data}^{(i)}$ corresponding to $\theta^{(i)}$. If the generated data set $\text{Data}^{(i)}$ constitutes a good realization as defined above, we keep $\theta^{(i)}$ as a true parameter value to be estimated and treat the generated data $\{S_t^{(i)}, I_t^{(i)}, R_t^{(i)}\}$ as observed data. We repeat the process until we obtain 100 underlying true parameter values $\theta^{(i)}$ and the corresponding 100 datasets $\{S_t^{(i)}, I_t^{(i)}, R_t^{(i)}\}$. For simplicity, we fix the initial infected node at node 1 and set the simulation time period $T = 100$ for all $i$.

\textit{Step 2. Estimating parameters}: For each iteration $i$, $i \in \{1,\cdots,100\}$, based on the sequence of $\{S_t^{(i)}, I_t^{(i)}, R_t^{(i)}\}$, we use RABC to estimate the underlying true parameter value $\theta^{(i)} $. In this estimation step, we chose priors for $\beta$ as $ U(0,1) $, $\gamma$ as $U(0, .5) $, the final threshold as 40, and sampled 100 particles to form the posterior. We also used the simple Euclidean distance, $\mbox{D}= \sqrt{ \sum_{t=1}^{99} (I(t)-I^{(s)}(t))^2}+ $ $\sqrt{ \sum_{t=1}^{99} ((R(t)-R^{(s)}(t))^2}$, where $t=1,\ldots,99$ are the days during the study period, $I(t)$ is the number of  infected nodes and $R(t)$ is the number of recovered nodes at time $t$; $I^{(s)}(t)$ and $R^{(s)}(t)$ are the corresponding numbers from simulated data.

\textit{Step 3. Evaluating parameter estimates}: For each synthesized data set $i$, $i \in \{1,\cdots,100\}$, we evaluated the accuracy of our parameter estimates for each method based on coverage rate of the interquartile  (IQ Cover), coverage rate of the 95 percentile interval ($95 \%$ Cover), and the interquartile range (IQR), for each parameter $ \beta^{(i)}, \gamma^{(i)}, \gamma^{(i)}$. We also compared the average time requirements to obtain the estimators corresponding to each realization using each method. The computation time is based on the results after submitting the parallel Python code to the University of Tennessee of Chattanooga Cluster with 3GB of memory and one CPU per task.  

Table \ref{accuracy2methods} shows that the average transmission matrix model estimated model parameters roughly as accurately as the proposed models, but it significantly reduced the computation time (down from 13.7 hours to 0.4 hours). This was to be expected as the average transmission matrix model requires network information in order to obtain the transmission matrix; once the transmission matrix is available, the spreading process can proceed at the same rate as the mass-action model. This important aspect addresses a significant challenge associated with the utilization of ABC in network infectious disease epidemiology research: the lengthy computational time required to directly simulate epidemic data on network for calibration purposes. The average transmission matrix approach is thus an excellent candidate for implementing ABC in network epidemiology.

\begin{table}

\captionof{table}{Comparison of parameter estimation between the proposed SIR model and the SIR ATMM using RABC. 
\label{accuracy2methods}}
\centering

\begin{tabular}{lrrrrr}

  \hline
  
Method & Average time (h)& Parameter & IQ Cover & $95 \%$ Cover& IQR    \\

  \hline

Proposed SIR & 13.7 & $\beta$ & 0.474 &   0.982 &   0.043     \\ 

&  &   $\gamma$ &   0.684 &   1.000 &   0.028 \\ 
  
 \hline

SIR ATMM  & 0.4  & $\beta$ & 0.456 &   0.965 &   0.042   \\ 

&  &   $\gamma$ &  0.702 &   1.000 &   0.027  \\ 
  
 \hline

\end{tabular}

\end{table}


\subsection{The mass-action model and the ATMM}
\label{ModifiedSIRVs.Mass-actionSIR}

In this section, we address a fundamental question: If network information were readily accessible, how useful would it be compared with merely using the mass-action model? We provide a quantitative answer to this question by comparing the fit of the average transmission matrix model and the mass-action model to  observed data, as well as the computation time required for each method. Specifically, we initially utilized ABC to estimate the model parameters for each model based on the observed data. We then used the ABC posteriors of model parameters to find the average realization and the 95\% confidence band for the number of current and cumulative infected cases (infected and recovered) for every approach.

We calculated the 95\% confidence interval for each approach as follows. We simulated three distinct SIR data sets using each model parameter sample of the ABC posteriors, and we retained only the best 30 simulated data that were closest to the observed data. The point of simulating three data sets for each model parameter is to avoid losing the particle (posterior sample) by chance, as the spreading process on a network might cause the realization to stop abruptly if the recovered nodes are in bottleneck positions at the early stages of the spreading process. From the best 30 realizations, we constructed a 95\% confidence interval for each method. In addition, for each of the 30 realizations, we calculated the Euclidean distance as defined in Section \ref{ModifiedSIRvsProposedSIR}. Based on these distances, we calculated the mean distance and its standard deviation.  

\begin{figure}[h]
     \centering\includegraphics[width=\textwidth]{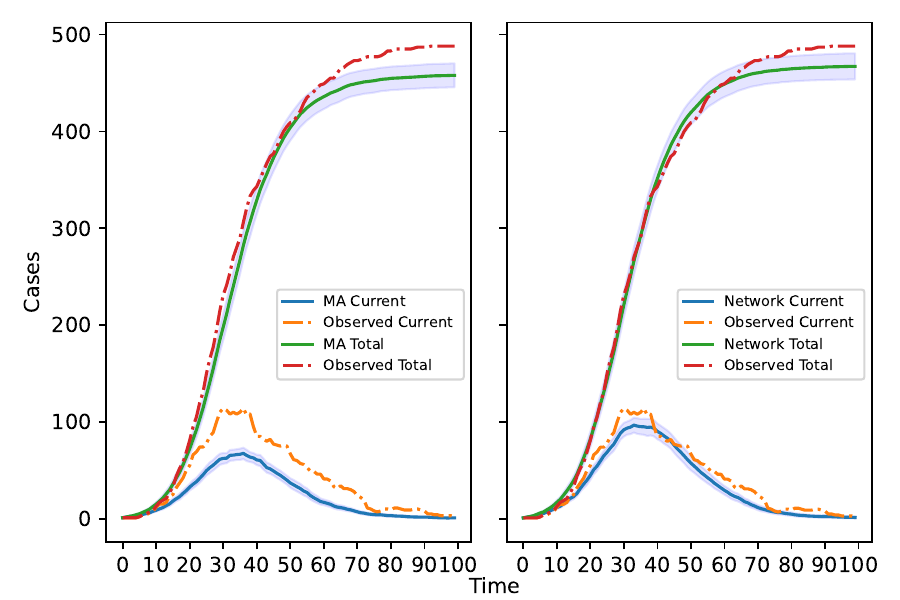}
         \caption{Comparison of the 95\% confidence band for the mass-action model and the ATMM.}
         \label{NaivevsNetworkCI}
\end{figure}

Figure \ref{NaivevsNetworkCI} demonstrates that network information provides a far better fit to the observed data. The 95\% confidence band derived from the average transmission matrix method effectively captures the observed data. However, when naively applying the mass-action model to fit the spreading process on the network, the results deviate significantly from the observed data. Table \ref{maVSnetwork} provides more information on the distance and time for each model. Here, we take the final threshold as 40 for the ABC procedure. The table shows that adopting the mass-action model to match the epidemic data naively not only results in a worse fit than the average transmission matrix model but also requires a threefold increase in processing time. This interesting phenomenon arises because the computer is having difficulty finding a suitable fit between data generated by the mass-action model and the observed network data. If the ABC acceptance threshold is lower than 40, the mass-action model will eventually fail to converge because we are using a wrong model to fit the network epidemic.  Therefore, network knowledge is extremely valuable and can provide insights into the nature of epidemics.

\begin{table}

\captionof{table}{Comparison of the mass-action and the naive method. 
\label{maVSnetwork}}
\centering

\begin{tabular}{lrrrrrr}

  \hline
  
Method & Time (h)  & Mean Distance   & Std Distance    \\

  \hline

Mass-action & 2.1 &  46.593 &   2.770 \\

 ATMM & 0.7 &  44.566 &   1.783 \\

 \hline

\end{tabular}

\end{table}

\section{Discussion}\label{discussion}
In this study, we examined the connection between network models and mass-action models. We proposed a spreading rule on networks that allows for an exact match between the epidemic spread on the network and the classic mass-action models when the graph is fully connected. We then developed modified spreading processes on networks that are similar to the classic mass-action models. We also proved that the modified processes and the proposed spreading rule on networks have the same average number of infections. Our results also reveal the differences between the two models as well as how the network model differs from the traditional mass-action model. Besides considering the popular SI and SIR spreading processes on networks, we extended the SITAD model to networks. We also analyzed and compared outbreaks during the early stage of the SIR spreading process for network and mass-action models. Ultimately, by utilizing the synthesized data from an empirical network, we highlighted the benefits of the proposed method.

The main limitation of the modified process is that the network is fixed. In practice, the network may evolve over time. There are also many extensions that can be conducted using the proposed spreading process, such as investigating different epidemiological quantities such as the basic reproduction number $R_0$, the exact/approximate solution of the spreading process, and  prevention strategies on the network. Finally, as the community focuses more on network epidemiology to better understand the dynamics of disease spread, particularly sexually transmitted diseases, we anticipate that tools and methods from network science will be increasingly used to investigate disease prevention and control strategies.

\section*{Acknowledgments}
This research was supported by the NIH award AI138901 (Onnela). The authors thank to Dr. Louis Raynal for useful insights and for sharing the RABC Python code.

\section*{Contributions} 
T.M.L. and J.P.O. designed the research; T.M.L. performed the research. J.P.O supervised the research; T.M.L. and J.P.O. wrote and edited the paper.

\section*{Change of Institution}
 T.M.L. started the project at Harvard Chan School of Public Health, Harvard University,
Boston, Massachusetts, U.S.A.

\section*{Data and materials availability}
The empirical network data is publicly available at The Copenhagen Networks Study interaction data. \cite{sapiezynski2019interaction} Python code used in this study is publicly available at https://github.com/onnela-lab/connecting-ma-network.

\section*{Appendix}
\label{appendix}
The appendix includes the proofs of lemmas in the main text and pseudo-code for some commonly used network spreading algorithms, as well as the proposed SI, SIR, SITAD spreading rules and the sampling process for determining infection order sequence.

\noindent \textit{Proof of Lemma \ref{SIlemma}:} Lemma \ref{SIlemma} is a direct result from Lemma \ref{SIRlemma}. Notice that if we replace the Poisson distribution with the Binomial distribution, the number of infections of the two approaches give rise to the same distribution. \hfill{Q.E.D.}

\noindent \textit{Proof of Lemma \ref{SIRlemma}:} 

We need to prove that 
\begin{equation}
\label{inductionSIR}
   E\big(  I_{\bold{T},k}^+\big) = E\big(  I_k^+\big), E\big(  R_{\bold{T},k}^+ \big) = E\big(  R_k^+ \big) 
\end{equation} for all $k = 1, 2, \cdots$. We use induction to prove (\ref{inductionSIR}). For $k = 1$, on a given transmission matrix $\bold{T}$, we have 
$ E\big(  I_{\bold{T},1}^+ \big) =  E\big( I_{\bold{T},1}\vert I_{\bold{T},0} = 1, R_{\bold{T},0} = 0  \big) =  E \big( I_\bold{T}(0) \big) + E\big(  I_\bold{T}(0) \big) \beta \bar{T}_1 - \gamma I_\bold{T}(0)  = 1 + \beta \bar{T}_1 - \gamma$, where $\bar{T}_1 = \sum_{i=1}^N \bold{T}[1,i]/1 = \sum_{i=1}^N \bold{T}[1,i]$. On the other hand, the transmission on the network gives us: 
$E\big(  I^+(1) \big) = E\big(  I(1)\vert I(0) = 1, R(0) = 0  \big) = 1 + \sum_{j \in I(0)} S_j/(N_j + 1) - \gamma   $.
Since the infection order is known, $\sum_{j \in I(0)} S_j/(N_j + 1) = \sum_{i=1}^N \bold{T}[1,i] = \bar{T}_1$. Therefore, $ E \big(I^+(1)\big) = E\big(  I_\bold{T}^+(1) \big)$. Suppose that (\ref{inductionSIR}) holds up to $k$, i.e., $ E \big(I^+(k)\big) = E\big(  I_\bold{T}^+(k) \big), E \big(R^+(k)\big) = E\big(  R_\bold{T}^+(k) \big)$, we need to prove that (\ref{inductionSIR}) also hold for $(k+1)$, i.e., we need to prove that $ E \big(I^+(k+1)\big) = E\big(  I_\bold{T}^+(k+1) \big), E \big(R^+(k+1)\big) = E\big(  R_\bold{T}^+(k+1) \big)$. We have  $E \big( R_\bold{T}^+(k+1) \big) = E\big(  R_\bold{T}^+(k) \big) + \gamma E\big( I_\bold{T}^+(k) \big)$, $E \big( R^+(k+1) \big) = E\big(  R^+(k) \big) + \gamma E\big( I_\bold{T}^+(k) \big)$. From the induction hypothesis, we have $E\big(  R_\bold{T}^+(k) \big) =E\big(  R^+(k) \big) $ and $E\big( I_\bold{T}^+(k) \big) = E\big( I_\bold{T}^+(k) \big)$, therefore $E \big( R_\bold{T}^+(k+1) \big) = E \big( R^+(k+1) \big)$. Finally, we need to prove that $E\big(  I_\bold{T}^+(k+1) \big)  = E\big(  I^+(k+1) \big)$. Since  $E\big(  I_\bold{T}^+(k+1) \big) = E\big( I_\bold{T}(k+1)\vert I_\bold{T}(k) = E\big(  I_\bold{T}^+(k) \big), R_\bold{T}(k) = E\big(R_\bold{T}^+(k) \big)  \big) = E\big(  I_\bold{T}^+(k) \big) + E\big(  I_\bold{T}^+(k) \big) \beta \bar{T}_{H_\bold{T}(k)} -  \gamma E\big(  I_\bold{T}^+(k) \big) $, where $H_\bold{T}(k) = E\big(  I_\bold{T}^+(k) + R_\bold{T}^+(k) \big)$. On the other hand, we have $E\big(  I^+(k+1) \big) = E\big( I(k+1)\vert I(k) = E\big(  I^+(k) \big), R(k) = E\big(  R^+(k) \big) \big) = E \big(  I^+(k) \big) + E\big(  I^+(k) \big) \beta  \big(\sum_{j \in \bold{H}(k)} S_j/(N_j+1) \big)/ H(k) - \gamma E \big(  I^+(k) \big)$ ,  where $\bold{H}(k)$ is the set of all currently infected nodes and $H(k) = E\big(  I^+(k) + R^+(k) \big)$. By the induction hypothesis, we have $E\big(  I^+(k)  \big) = E\big(  I_\bold{T}^+(k)  \big), E\big(   R^+(k) \big) = E\big(   R_\bold{T}^+(k) \big)$, therefore $H(k) = H_\bold{T}(k) $. Since the infection order is known, $\sum_{j \in \bold{H}(k)} S_j/(N_j+1) = H(k) \bar{T}_{H(k)} = H(k) \bar{T}_{H_\bold{T}(k)}$. So, $E\big(  I^+(k+1) \big) = E \big(  I^+(k) \big) + E\big(  I^+(k) \big) \beta  \bar{T}_{H_\bold{T}(k)} - \gamma E \big(  I^+(k) \big) = E \big(  I_\bold{T}^+(k) \big) + E\big(  I_\bold{T}^+(k) \big) \beta  \bar{T}_{H_\bold{T}(k)} - \gamma E \big(  I_\bold{T}^+(k) \big) = E\big(  I_\bold{T}^+(k+1) \big)$. The lemma is proved. 

\noindent \textit{Proof of Lemma \ref{SITADlemma}:} Repeat the argument as in Lemma \ref{SIRlemma}.

\noindent \textit{Proof of Proposition \ref{Rt_agressive}}:
\begin{itemize}
    \item[a.] The effective reproductive number for the mass-action model when it has $h$ infected nodes is $R^C_t = \dfrac{\beta}{ \gamma} (1 - \dfrac{h}{N}) \xrightarrow{N \rightarrow \infty} \dfrac{\beta}{ \gamma}$. We also have $R^U_t = \dfrac{\beta}{ \gamma} (1 - \dfrac{2}{h} \sum_{i \in \bold{I}}  \dfrac{1}{k_i + 1}) = \leq  \dfrac{\beta}{ \gamma} (1 - \dfrac{2}{N}) \xrightarrow{N \rightarrow \infty} \dfrac{\beta}{ \gamma}$. Therefore, for large network with $N \rightarrow \infty$, at the early stage, the effective reproductive number $R_t$ on the network is always asymptotically bounded by the effective reproductive number of the mass action model.
\item[b.] The upper bound and lower bound of $R^U_t$ are given by $ \dfrac{\beta}{ \gamma} (1 - \dfrac{2}{k^*_\bold{I}  + 1})  \leq R^U_t \leq  \dfrac{\beta}{ \gamma} (1 - \dfrac{2}{N}).$ We also observe that that $ R^C_t = \dfrac{\beta}{ \gamma} (1 - \dfrac{h}{N}) \leq  \dfrac{\beta}{ \gamma} (1 - \dfrac{2}{k^*_\bold{I}  + 1})$ iff $ k^*_\bold{I} \geq \dfrac{2N}{h} - 1.$ So, for finite networks of size $N$ with $h$ infected nodes, if the spreading process of $h$ nodes occurred in a line graph and each infected node has more than $\dfrac{2N}{h} - 1$ neighbors, the epidemic on the network at that time is more aggressive than the epidemic on the mass-action model. 
\end{itemize}

Algorithms \ref{GillispieSI}-\ref{samplingorder} below provide the pseudo-codes for different spreading rules and procedures mentioned in the main text. More specifically, Algorithms \ref{GillispieSI}-\ref{degreeinfectivity} are the pseudo-code for the commonly spreading rules used in practice, including the Gillespie, the unit infectivity, and the degree infectivity; Algorithms \ref{SIproposed}-\ref{SITADnetwork} are the pseudo-codes for the proposed spreading rules for the SI, SIR, and SITAD processes; and Algorithm \ref{samplingorder} is the pseudo-code for the sampling procedure to obtain the infection order sequence. 


\begin{algorithm}[t!]
\caption{The Gillespie SI spreading rule}\label{GillispieSI}
\begin{algorithmic}
\State \textbf{Input:} Network $G$, per-edge transmission rate $\tau$, initial\_infections, maximum time $T$.

\State \textbf{Output:} List times, number of $S, I$ at each time \\

\State \quad  \textbf{function} GillespieSI\_network\_spread($G, \tau, T$, initial\_infections)

\State \quad \quad  time, $S, I \gets$ [0], [$\vert G \vert$ - len(initial\_infections)], [len(initial\_infections)]

\State \quad \quad  infected\_nodes $\gets$ initial\_infections

\State \quad \quad  at\_risk\_nodes $\gets$ susceptible neighbors of infected

\State   \quad \quad  \textbf{for} each node $u$ in at\_risk\_nodes \textbf{do}

\State \quad \quad \quad \quad infection\_rate$[u]$ = $\tau \times$ number of infected neighbors

\State \quad \quad 
total\_infection\_rate = $\sum_{u \in \text{at\_risk\_nodes}}$infection\_rate$[u]$

\State \quad \quad 
time $\gets$ exponential\_variate(total\_infection\_rate)

\State \quad \quad \textbf{while} time $< T$ \textbf{do}

\State \quad \quad \quad \quad choose $u$ from at\_risk\_nodes with probability $p_u$ ( $p_u$ = infection\_rate$[u]$/total\_infection\_rate)

\State \quad \quad \quad \quad remove $u$ from at\_risk\_nodes

\State \quad \quad \quad \quad add $u$ to infected\_nodes

\State \quad \quad \quad \quad \textbf{for} susceptible neighbors $v$ of u \textbf{do}

\State \quad \quad \quad \quad \quad \textbf{if} $v$ not in at\_risk\_nodes \textbf{then}

\State \quad \quad \quad \quad \quad \quad add $v$ in at\_risk\_nodes

\State \quad \quad \quad \quad \quad update infection\_rate$[v]$

\State \quad \quad \quad \quad  update times, $S, I$

\State \quad \quad \quad \quad
time $\gets$ time + exponential\_variate(total\_infection\_rate)

\State \quad \quad \quad
\textbf{return} times, $S, I$

\end{algorithmic}
\end{algorithm}


\begin{algorithm}[t!]
\caption{The unit infectivity SI spreading rule}\label{unitinfectivity}
\begin{algorithmic}
\State \textbf{Input:} Network $G$, per-edge transmission rate $\tau$, initial\_infections, maximum time $T$.

\State \textbf{Output:} Number of $S, I$ at each time \\

\State \quad  \textbf{function} UnitInfectivitySI\_network\_spread($G, \tau, T$, initial\_infections)

\State \quad \quad  time, $S, I \gets$ 0,  [$\vert G \vert$ - len(initial\_infections)], [len(initial\_infections)]

\State \quad \quad  infected\_nodes $\gets$ initial\_infections

\State   \quad \quad  \textbf{while} time $<T$ \textbf{do}

\State   \quad \quad \quad \textbf{for} each node $u$ in infected\_nodes \textbf{do}

\State   \quad \quad \quad \quad randomly choose a neighbor $v$ of $u$

\State   \quad \quad  \quad \quad \textbf{if} $v$ is susceptible \textbf{then}

\State  \quad \quad \quad \quad  \quad \quad \quad add $v$ to infected\_nodes with probability $\tau$

\State  \quad \quad \quad \quad  \quad \quad \quad update S, I

\State  \quad \quad \quad time $\gets$ time+ 1

\State  \quad \quad \textbf{return} S, I
\end{algorithmic}
\end{algorithm}


\begin{algorithm}[t!]
\caption{The degree infectivity SI spreading rule}\label{degreeinfectivity}
\begin{algorithmic}
\State \textbf{Input:} Network $G$, per-edge transmission rate $\tau$, initial\_infections, maximum time $T$.

\State \textbf{Output:} Number of $S, I$ at each time \\

\State \quad  \textbf{function} DegreeInfectivitySI\_network\_spread($G, \tau, T$, initial\_infections)

\State \quad \quad  time, $S, I \gets$ 0,  [$\vert G \vert$ - len(initial\_infections)], [len(initial\_infections)]

\State \quad \quad  infected\_nodes $\gets$ initial\_infections

\State   \quad \quad  \textbf{while} time $<T$ \textbf{do}

\State   \quad \quad \quad \textbf{for} each node $u$ in infected\_nodes \textbf{do}

\State   \quad \quad \quad \quad \textbf{for} each neighbor $v$ of $u$ \textbf{do}

\State   \quad \quad  \quad \quad \quad \textbf{if} $v$ is susceptible \textbf{then}

\State  \quad \quad \quad \quad \quad  \quad \quad \quad add $v$ to infected\_nodes with probability $\tau$

\State  \quad \quad \quad  update S, I

\State  \quad \quad \quad time $\gets$ time + 1 

\State  \quad \quad \textbf{return} S, I
\end{algorithmic}
\end{algorithm}


\begin{algorithm}[t!]
\caption{The proposed SI spreading rule on networks}\label{SIproposed}
\begin{algorithmic}
\State \textbf{Input:} Network $G$, transmission rate $\beta$, initial\_infections, maximum time $T$.

\State \textbf{Output:} List  number of $S, I$ at each time \\

\State \quad  \textbf{function} ProposedSI\_network\_spread($G, \beta, T$, initial\_infections)

\State \quad \quad  time, $S, I \gets$ 0, [$\vert G \vert$ - len(initial\_infections)], [len(initial\_infections)]

\State \quad \quad  infected\_nodes $\gets$ initial\_infections

\State \quad \quad  at\_risk\_nodes $\gets$ susceptible neighbors of infected

\State \quad \quad \textbf{while} time $< T$ \textbf{do}

\State   \quad \quad \quad  \textbf{for} each node $u$ in at\_risk\_nodes \textbf{do}

\State \quad \quad \quad \quad \quad update weights W$[u]$ = number infected neighbors of $u$

\State   \quad \quad \quad  \textbf{for} each node $v$ in infected\_nodes \textbf{do}

\State   \quad \quad \quad \quad transmission\_rate$[v]$ = susceptible neighbors of $v$/(neighbors of $v$ + 1)

\State   \quad \quad \quad \quad at\_risk\_nodes $\gets$ at\_risk\_nodes + susceptible neighbors of $v$

\State   \quad \quad \quad  total\_transmission\_rate = $\beta \times$ sum(transmission\_rate) 

\State \quad \quad \quad  new\_infections $\gets$ Binomial\_variate(at\_risk\_nodes, total\_transmission\_rate/ at\_risk\_nodes)

\State \quad  \quad \quad \textbf{if} new\_infections $> 0$ \textbf{do}

\State \quad  \quad \quad \quad choose new\_infections nodes from at\_risk\_nodes by sampling with weight $W$ 

\State \quad  \quad \quad update  susceptible\_nodes, infected\_nodesat\_risk\_nodes, S, I

\State \quad  \quad \quad
time $\gets$ time + 1

\State \quad \quad
\textbf{return} times, $S, I$

\end{algorithmic}
\end{algorithm}


\begin{algorithm}[t!]
\caption{The proposed SIR spreading rule on networks}\label{SIRnetwork}
\begin{algorithmic}
\State \textbf{Input:} Network $G$, model parameter $\bold{\theta} = (\beta, \gamma)$, initial\_infections, maximum time $T$.

\State \textbf{Output:} List  number of $S, I, R$ at each time \\

\State \quad  \textbf{function} ProposedSIR\_network\_spread($G, \bold{\theta}, T$, initial\_infections)

\State \quad \quad  time, $S, I, R \gets$ 0, [$\vert G \vert$ - len(initial\_infections)], [len(initial\_infections)], [0]

\State \quad \quad  infected\_nodes $\gets$ initial\_infections

\State \quad \quad  at\_risk\_nodes $\gets$ susceptible neighbors of infected
\State \quad \quad  recover\_nodes $\gets$ []

\State \quad \quad \textbf{while} time $< T$ \textbf{do}

\State   \quad \quad \quad  \textbf{for} each node $u$ in at\_risk\_nodes \textbf{do}

\State \quad \quad \quad \quad \quad update weights W$[u]$ = number infected neighbors of $u$

\State   \quad \quad \quad  \textbf{for} each node $v$ in infected\_nodes \textbf{do}

\State   \quad \quad \quad \quad transmission\_rate$[v]$ = susceptible neighbors of $v$/(neighbors of $v$ + 1)

\State   \quad \quad \quad \quad at\_risk\_nodes $\gets$ at\_risk\_nodes + susceptible neighbors of $v$

\State   \quad \quad \quad  total\_transmission\_rate = $\beta \times$ sum(transmission\_rate) 

\State \quad \quad \quad  new\_infections $\gets$ Binomial\_variate(at\_risk\_nodes, total\_transmission\_rate/ at\_risk\_nodes)

\State \quad \quad \quad  new\_recovers $\gets$ Binomial\_variate(number of infected\_nodes, $\gamma$)

\State \quad  \quad \quad \textbf{if} new\_infections $> 0$ \textbf{do}

\State \quad  \quad \quad \quad choose new\_infections nodes from at\_risk\_nodes by sampling with weight $W$ 

\State \quad \quad \quad  randomly choose new\_recovers nodes from infected\_nodes

\State \quad  \quad \quad update  susceptible\_nodes, infected\_nodes, recover\_nodes, at\_risk\_nodes, S, I, R

\State \quad  \quad \quad
time $\gets$ time + 1

\State \quad \quad
\textbf{return} times, $S, I, R$

\end{algorithmic}
\end{algorithm}


\begin{algorithm}[t!]
\caption{The proposed SITAD spreading rule on networks}\label{SITADnetwork}
\begin{algorithmic}
\State \textbf{Input:} Network $G$, model parameter $\bold{\theta} = (\beta_1, \beta_2, \gamma_1, \delta_1, \gamma_2, \delta_2)$, initial\_HIV, maximum time $T$.

\State \textbf{Output:} List  number of $S, I, T, A, D$ at each time \\

\State \quad  \textbf{function} ProposedSITAD\_network\_spread($G, \bold{\theta}, T$, initial\_HIV)

\State \quad \quad  time, $S, I, T, A, D \gets$ 0, [$\vert G \vert$ - len(initial\_HIV)], [len(initial\_HIV)], [0], [0], [0]

\State \quad \quad  HIV\_nodes $\gets$ initial\_HIV

\State \quad \quad  treated\_nodes $\gets$ []
\State \quad \quad  AIDS\_nodes $\gets$ []
\State \quad \quad  death\_nodes $\gets$ []

\State \quad \quad  at\_risk\_nodes $\gets$ susceptible neighbors of HIV\_nodes and AIDS\_nodes

\State \quad \quad \textbf{while} time $< T$ \textbf{do}

\State   \quad \quad \quad  \textbf{for} each node $u$ in at\_risk\_nodes \textbf{do}

\State \quad \quad \quad \quad \quad update weights W$[u]$ = $\beta_1 \times$ number HIV neighbors of $u$ + $\beta_2 \times$ number AIDS neighbors of $u$

\State   \quad \quad \quad  \textbf{for} each node $v_1$ in HIV\_nodes \textbf{do}

\State   \quad \quad \quad \quad HIV\_rate$[v_1]$ = susceptible neighbors of $v_1$/(neighbors of $v_1$ + 1)
\State   \quad \quad \quad \quad at\_risk\_nodes $\gets$ at\_risk\_nodes + susceptible neighbors of $v_1$

\State   \quad \quad \quad  \textbf{for} each node $v_2$ in AIDS\_nodes \textbf{do}

\State   \quad \quad \quad \quad AIDS\_rate$[v_2]$ = susceptible neighbors of $v_2$/(neighbors of $v_2$ + 1)

\State   \quad \quad \quad \quad at\_risk\_nodes $\gets$ at\_risk\_nodes + susceptible neighbors of $v_2$

\State   \quad \quad \quad  total\_transmission\_rate = $\beta_1 \times$ sum(HIV\_rate) + $\beta_2 \times$ sum(AIDS\_rate) 

\State \quad \quad \quad  new\_HIV $\gets$ Binomial\_variate(at\_risk\_nodes, total\_transmission\_rate/ at\_risk\_nodes)

\State \quad \quad \quad  HIV\_treated $\gets$ Binomial\_variate(number of HIV\_nodes, $\gamma_1$)

\State \quad \quad \quad  AIDS\_treated $\gets$  Binomial\_variate(number of AIDS\_nodes, $\gamma_2$)

\State \quad \quad \quad  new\_treated $\gets$ HIV\_treated + AIDS\_treated

\State \quad  \quad \quad \textbf{if} new\_HIV $> 0$ \textbf{do}

\State \quad  \quad \quad \quad choose new\_HIV nodes from at\_risk\_nodes by sampling with weight $W$ 

\State \quad \quad \quad  randomly choose HIV\_treated nodes from HIV\_nodes

\State \quad \quad \quad  randomly choose AIDS\_treated nodes from AIDS\_nodes

\State \quad \quad \quad  new\_AIDS $\gets$  Binomial\_variate(number of HIV\_nodes, $\delta_1$)

\State \quad \quad \quad  new\_death $\gets$  Binomial\_variate(number of AIDS\_nodes, $\delta_2$)

\State \quad \quad \quad  randomly choose new\_AIDS nodes from HIV\_nodes

\State \quad \quad \quad  randomly choose new\_death nodes from AIDS\_nodes

\State \quad  \quad \quad update  susceptible\_nodes, HIV\_nodes, treated\_nodes, AIDS\_nodes, death\_nodes,  at\_risk\_nodes, S, I, T, A, D

\State \quad  \quad \quad
time $\gets$ time + 1

\State \quad \quad
\textbf{return} times, $S, I, T, A, D$

\end{algorithmic}
\end{algorithm}


\begin{algorithm}[t!]
\caption{The sampling mechanism to obtain the infection order sequence   }\label{samplingorder}
\begin{algorithmic}
\State \textbf{Input:} Network $G$, initial\_infections

\State \textbf{Output:} The infection order sequence \\

\State \quad  \textbf{function} Infection\_order($G$, itial\_infections)

\State \quad \quad  infected\_nodes, susceptible\_nodes $\gets$ initial\_infections, $G \setminus$  initial\_infections

\State \quad \quad  at\_risk\_nodes $\gets$ susceptible neighbors of infected

\State   \quad \quad \quad  \textbf{for} each node $u$ in at\_risk\_nodes \textbf{do}

\State \quad \quad \quad \quad \quad update weights W$[u]$ = number of infected neighbors

\State \quad  \quad \quad \quad choose one new infected node from at\_risk\_nodes by sampling with weight $W$ 

\State \quad  \quad \quad update  susceptible\_nodes, infected\_nodes, infected order sequence

\State \quad \quad
\textbf{return} infected order sequence

\end{algorithmic}
\end{algorithm}

\clearpage

\bibliographystyle{unsrt}

\end{document}